\documentclass[a4paper,11pt]{article}
\pdfoutput=1 

\usepackage{jheppub} 

\usepackage{comment}
\usepackage{amsmath,amssymb,mathtools,graphicx,setspace}
\usepackage{slashed} 		 
\usepackage{color}
\usepackage{caption}
\makeatother
\newcommand{\be}{\begin{equation}}
\newcommand{\bea}{\begin{eqnarray}}
\newcommand{\eea}{\end{eqnarray}}
\newcommand{\ba}{\begin{array}}
\newcommand{\ea}{\end{array}}
\newcommand{\ee}{\end{equation}}
\newcommand{\bes}{\begin{equation*}}
\newcommand{\beas}{\begin{eqnarray*}}
\newcommand{\eeas}{\end{eqnarray*}}
\newcommand{\bas}{\begin{array*}}
\newcommand{\eas}{\end{array*}}
\newcommand{\ees}{\end{equation*}}

\def\half{{1 \over 2 }}

\title{On the saturation of late-time growth of complexity in supersymmetric JT gravity}

\author[a]{Mohsen Alishahiha}
\emailAdd{alishah@ipm.ir}
\affiliation[a]{School of Physics, Institute for Research in Fundamental Sciences (IPM),\\
	P.O. Box 19395-5531, Tehran, Iran\\} 
\author[b,c]{and Souvik Banerjee}
\emailAdd{souvik.banerjee@physik.uni-wuerzburg.de}
\affiliation[b]{Institut f{\"u}r Theoretische Physik und Astrophysik,
	Julius-Maximilians-Universit{\"a}t W{\"u}rzburg,\\ Am Hubland, 97074 W{\"u}rzburg, Germany\\} 
\affiliation[c]{W\"urzburg-Dresden Cluster of Excellence ct.qmat\\}

\abstract{In this work we use the modified replica trick, proposed in arXiv:2205.01150, to compute the late time behaviour of complexity for JT gravity with ${\cal N} = 1$ and ${\cal N} = 2$ supersymmetries. For the ${\cal N} = 1$ theory, we compute the late time behaviour of complexity defined by the ``quenched geodesic length" and obtain the expected saturation of complexity at time $t \sim e^{S_0}$, to a constant value with time-independent variance. For the ${\cal N} = 2$ theory, we explicitly compute complexity at the disk level which yields the late-time linear growth of complexity. However, we comment on the expectation of the late-time saturation by speculating the trumpet partition function and the non-perturbative corrections to the spectral correlation, relevant for the late time behaviour of complexity. Furthermore, we compute the matter correlation functions for both the theories.}

\begin{document} 
\maketitle
\flushbottom

\section{ Introduction}
In the light of AdS/CFT correspondence \cite{Maldacena:1997re, Witten:1998qj,Gubser:1998bc}, the quantum computational complexity has been thought of as one of the most useful probes to study the physics of black hole interior. This is a curious quantity which, for a finite entropic fast-scrambling system, keeps growing long after the system attains thermal equilibrium \cite{Susskind:2014rva,Susskind:2018pmk}. Therefore, in a holographic setting, it is naturally identified with the growth of the black hole interior. Accordingly, the most natural holographic proposal to compute complexity in AdS/CFT turns out to be the celebrated ``Complexity = Volume" (CV) conjecture which measures complexity as the volume of the maximal slice in the interior of the black hole \cite{Susskind:2014rva, Stanford:2014jda}.

For a typical chaotic Hamiltonian, complexity is expected to grow linearly up to time $t \sim \left( \mathcal{O}(e^S)\right)$, following which
one expects a saturation to a plateau at later times. The late time value of complexity is expected to be of ${\mathcal{O}}(e^S)$, $S$ being the total entropy of the system \cite{Susskind:2015toa,Brown:2016wib,Balasubramanian:2019wgd,
Susskind:2020wwe,Balasubramanian:2021mxo,Haferkamp:2021uxo}. However, describing this plateau region in the holographic setting has been a challenge for the last few years. To attack this problem, it was illustrative to study complexity in Jackiw-Teitelboim (JT) gravity which is a theory of two-dimensional dilaton gravity
with the Lorentzian action 
\cite{Jackiw:1984je,Teitelboim:1983ux}
\begin{align}\label{eq:LorentzianJTaction}
    S_{\text{JT}}=\frac{S_0}{2 \pi}\left( \int \sqrt{-g}R + 2 \int \sqrt{|h|} K \right) + \int \sqrt{-g} \phi 
   \left(R+2\right) + 2 \int \sqrt{|h|} \phi \left(K-1\right)\,,
\end{align}
where the first term is the topological Gauss-Bonnet term and $S_0$ is the ground state entropy. 
The asymptotic boundary conditions are set by fixing the value of the induced metric 
and the dilaton on a UV cutoff near the AdS boundary \cite{Maldacena:2016upp,Jensen:2016pah,Engelsoy:2016xyb}.

The advantage of choosing this two dimensional model of gravity is that, in this model, we are in perfect control to study the Euclidean action in full glory including arbitrary 
genus and hyperbolic Riemann surfaces which amounts to computing exponentially small corrections to semi-classical gravity calculations \cite{Maldacena:2016upp,Engelsoy:2016xyb,Saad:2018bqo, Saad:2019lba}.
In particular, incorporating geometries with an arbitrary number of asymptotic boundaries and arbitrary genus in the gravitational path integral is shown to be dual to a specific double-scaled Hermitian matrix integral. Accordingly, the JT gravity models
exhibit typical spectral statistics at late times with a ``dip-ramp-plateau" structure of correlation functions 
\cite{Saad:2018bqo,Saad:2019lba,Saad:2019pqd,Altland:2020ccq}. Moreover, inclusion of higher 
topologies in the path integral has been shown to be instrumental to obtain a unitary Page curve \cite{Almheiri:2019qdq,Penington:2019kki}. While similar universal behaviour is also expected for higher dimensions \cite{Belin:2020hea,Cotler:2020ugk}, to establish an argument on universal behaviour of late time behaviour of complexity, two dimensional JT gravity models are the obvious starting point. 

The first study in this direction was in \cite{Iliesiu:2021ari} where complexity was defined in terms of a weighted sum over non-perturbative geodesic lengths involving non self-intersecting geodesics over arbitrary genus. The afore-mentioned weighted sum over geodesics was argued to be related to an analytically continued correlator of quantum field operator with the scaling dimension $\Delta$. More precisely, the complexity defined this way, turns out to be proportional to the derivative of this correlator with respect to $\Delta$ in the limit of $\Delta \rightarrow 0$.
This correlator incorporates the volume of the moduli space  of hyperbolic surfaces \cite{Mirzakhani:2006fta} through the weighted sum over an infinite number of geodesics. 

While this proposal yields the saturation of afore-defined complexity at late time, it is not satisfactory for two primary reasons. First, the limit $\Delta \rightarrow 0$ limit, used in the definition of the complexity is counter-intuitive from the notion of standard geodesic approximation employed for heavy operators in the AdS/CFT correspondence. Second, the computation of variance following this definition leads to the late time growth with time, which in turn puts a question mark over interpreting the weighted sum over geodesics as the complexity. Both these problems are satisfactorily resolved by introducing a modified replica trick \cite{Alishahiha:2022kzc}. Our formalism uses the Hartle-Hawking wavefunction constructed in \cite{Harlow:2018tqv} in the context of the factorization puzzle.

Given the Hartle-Hawking wavefunction $\Phi(\beta,\ell)$ corresponding  
to the integral over all Euclidean geometries with a given topology and the asymptotic AdS boundary of renormalised length $\beta$, one can compute the full quantum gravity 
expressions for the matter correlation functions in 
JT gravity  coupled to matter fields. Following \cite{Yang:2018gdb}, this procedure involves identifying a Kernel to the Hartle-Hawking wavefunction and using this to obtain the gravitational correlators through gravitational dressing of the corresponding quantum field theory correlation functions on fixed background AdS$_2$ .

While the computations of these correlation functions are straightforward with the knowledge of the Hartle-Hawking wavefuntion on a given topology, computation of complexity demands a little more work. This is where our modified replica trick comes handy. The holographic computation of complexity via the CV conjecture measures the volume of the Einstein-Rosen (ER) bridge which in two dimensions turns out to be the length of a geodesic extending between two points on the boundary. Using the notation where the classical geodesic distance is denoted by $\ell=- \ln y/2$,
the complexity is proportional to the expectation value of the geodesic ${\cal C}\sim \langle \ell\rangle_{\rm QG}$ in a theory of quantum gravity. It is worth stressing here that this expectation value does not involve any averaging over geodesics, rather, it denotes the effect of quantum gravitational dressing in a spirit similar to what we discussed above in the context of quantum gravity correlators above. Accordingly, the replica method that we discuss below does not assume any physical interpretation of summing over geodesics, unlike the one adopted in the work of \cite{Iliesiu:2021ari}.

Given the wavefunction $\Phi(\beta,\ell)$, the quantum expectation value of the geodesic length is given by
\be
\label{eq:replica-l-disk}
\langle \ell(u)\rangle =-\frac{1}{Z_{0}(\beta)}\int_0^\infty \frac{dy}{y}\;\Phi
(u,\ell) \;\Phi
(\beta-u,\ell)\;
(2\ln \frac{y}{4}).
\ee 
To evaluate this quantity, we use the modified replica trick \cite{Alishahiha:2022kzc} which involves expressing the logarithm in terms of the following limit\footnote{This is inspired by the replica trick used 
for computing the quenched free energy. See \cite{Engelhardt:2020qpv, Johnson:2020mwi, Johnson:2021rsh, Alishahiha:2020jko} for the computation of quenched free energy in JT gravity models.}
\be\label{RT}
\ln A=\lim_{N\rightarrow 0}\frac{A^N-1}{N}= \lim_{N\rightarrow 0}\frac{d}{dN}\;A^N\,.
\ee

We have normalised \eqref{eq:replica-l-disk} by ${Z_{0}^{-1}(\beta)}$, $Z_{0}(\beta)$ being the partition 
function on the given topology. Using the formula above, one can define complexity as
\be
\label{eq:replica-formula}
\langle \ell(u)\rangle =-\lim_{N\rightarrow 0}\frac{\langle y^{2N}\rangle_u-1}{N},
\ee
where
\be\label{yN}
\langle y^{2N}\rangle_u=\frac{1}{Z_{0}(\beta)}\int_0^\infty \frac{dy}{y}\;\Phi
(u,\ell) \;\Phi (\beta-u,\ell)\;\left( \frac{y}{4}\right)^{2N}
\ee
In order to obtain the late time behaviour of complexity, one needs to further analytically continue the Euclidean coordinate, $u$. It turns out that in order to get the late time saturation of complexity, one needs to go beyond the disk level and consider the connected trumpet contribution. The latter can be obtained from the knowledge of trumpet partition function and the form of the spectral correlation function including the non-perturbative effects. The modified replica trick described above smoothly sails through this case as well yielding the expected saturation of complexity as well as a time-independent variance at late time. In \cite{Alishahiha:2022kzc}, this computation was performed in presence of an end-of-the-world brane mimicking a one-sided black hole.

In the present work, we establish the applicability of the modified replica method to compute the late time behaviour of complexity for JT gravity with supersymmetry. For the ${\cal N} =1$ supersymmetric models, with a reasonable assumption, we achieve the expected saturation of complexity at late times $t \sim e^{S_0}$. 
We also show that just like the bosonic case, here also we find that at late times, the variance approaches a constant of the order of the square root of the complexity. For the ${\cal N} =2$ supersymmetric models, one can extend this formalism to compute the disk level complexity, which shows a linear growth of complexity at late times. We demonstrate this feature both for the left/ right sector as well as for the full Thermofield Double (TFD) state dual to a two-sided eternal black hole. However, extending this construction for the trumpet geometry is a bit tricky. This is because we do not know precisely the partition function and the spectral correlation function for these models. Nevertheless, using the intuitions coming from the computations of ${\cal N} =1$ and ${\cal N} = 0$ cases as well as the from disk level computation of ${\cal N} = 2$ case, it is possible to speculate the late time behaviour of complexity for JT gravity models with ${\cal N} = 2$ supersymmetry. With this, one can argue that the even for the JT gravity models with ${\cal N} = 2$ supersymmetry, one should expect a saturation of complexity at a late enough time. 

In this paper we also study the multiboundary partition functions and matter correlation functions for these models. We show that the matter correlation functions exhibit the dip-ramp-plateau
structure at late time, just like the spectral form factor. Furthermore we also comment on the correlation functions involving non-orientable topologies, for instance, the crosscap states considered in \cite{Yan:2022nod}. These states are important since they provide a late time decay of the correlators, even at the disk level.

\section{JT gravity with ${\cal N} = 1$ supersymmetry}
\label{sec:N1}
In this section we would like to study late times complexity growth for
the ${\cal N } =1$ super-Schwarzian theory. Different aspects of this model
have been analyzed in \cite{Fan:2021wsb}. For early works on this
subject see also \cite{Hikida:2007sz, Stanford:2017thb,Mertens:2017mtv}.
Most  of the ingredients we need for the computations to follow can be found in the papers mentioned above. Nevertheless, for completeness we will also review the necessary building blocks below.

The ${\cal N}=1$ model has $OSp(1|2)$ supergroup whose algebra is given by
\be
  [L_m, L_n] = (m-n)L_{m+n},\;\;\;\;
  \{F_r , F_s\} = 2L_{r+s},\;\;\;\;
  [L_m, F_r] = \left(\frac{m}{2}-r\right)  F_{m+r}\,.
\ee
The corresponding group element  can be written as
\be 
 g = e^{ - x_- L_-} e^{ \theta_- F_{-} } e^{  2\ell L_0 } e^{ \theta_+ F_{+} } e^{
 x_+ L_+}, 
 \ee 
where $x_\pm, \theta_\pm,\ell $ are the coordinates associated with each of the five generators. Using this notation, the representation of the left generators of the algebra, 
denoted by ${\cal L}_{\pm,0},{\cal F}_{\pm}$, can be written as follows.
\bea
 && {\cal L}_0 =-x_-\partial_{-} + \frac{1}{2}\partial_\ell - \half   \theta _-
 \hat \partial_{-} ,\;\;\;\;\;\; 	{\cal L}_- = -\partial_{-},\;\;\;\;\;
 {\cal F}_- =\hat\partial_{-} -  {\theta }_-\partial_{-} \nonumber\\
	&&
	{\cal L}_+ = e^{-2\ell }\partial_{+ }-x_-^2 \partial_{-} + x_- 
	\left( \partial_\ell  -  \theta _-\hat\partial_{-} \right)  
   - e^{ -\ell }        \theta_- ( \hat \partial_{+} + \theta_+ \partial_{+}) \nonumber\\ &&
 	{\cal F}_+ =e^{-\ell }(\hat\partial_{+} + \theta_+ \partial_{+}) + x_- 
 	 ( \hat\partial_{-} -  \theta_- \partial_{-} )    +    \theta_- \partial_\ell  \,
	\eea
	where $\partial_\pm=\partial_{x_\pm}$ and $\hat\partial_\pm=
	\partial_{\theta_\pm}$. There are also right generators satisfying an isomorphic algebra which we do not write here explicitly.
	The differential representation of the casimir operator is 
\be
{\cal C} = \frac{1}{4}\partial_\ell^2 + \frac{1}{4}  \partial_\ell + 
e^{ -2\ell} \partial_- \partial_+ - \frac{1}{2} e^{ -\ell}
 (\hat\partial_{-} + \theta_- \partial_-)
(\hat\partial_{+} + \theta_+ \partial_{+})
\ee

Form this expression of the casimir operator, one can write down the Schr\"odinger equation of the form \cite{Fan:2021wsb}
\be
\label{eq:SE-N1}
\left(-\partial_\ell\pm 2e^{-\ell}+4e^{-2\ell}\right)\psi_k(\ell)=k^2\psi_k(\ell),
\ee
which essentially describes a non-relativistic quantum particle
in a  Morse-like potential. 

Actually this equation can be also
obtained from  a super-Liouville quantum mechanics. The corresponding 
Lagrangian has three variables; one bosonic $\ell$ which could be interpreted as 
normalized geodesic  distance $\ell$ and two fermionic partners $\theta_\pm$.
This model has two wave functions 
obtained as the solutions\footnote{see \cite{Douglas:2003up, Fan:2021wsb} for more details about these solutions.} of the Schr\"odinger equation \eqref{eq:SE-N1}
\be
\label{eq:wavefunction-N1}
\Psi_{k,\pm}(\ell)=e^{-\ell/2}\left(K_{ik+\frac{1}{2}}(2e^{-\ell})
\pm K_{ik-\frac{1}{2}}(2e^{-\ell})\right)\,.
\ee

\subsection{The Hartle Hawking wavefunctions}

\begin{figure}
\begin{center}
\includegraphics[scale=1.2]{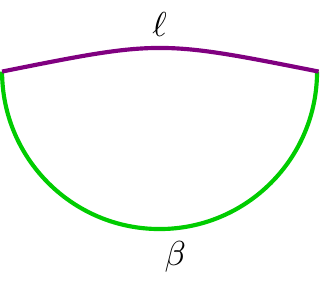}\hspace{3cm}
\includegraphics[scale=1.]{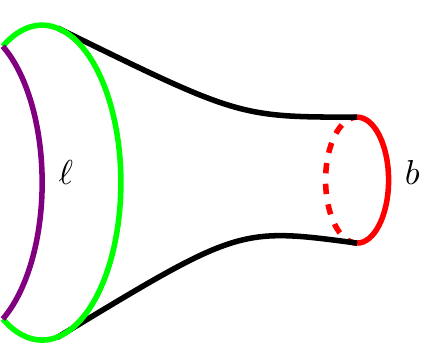}
\end{center}
\caption{ {\it Left:} The wavefunction of the Hartle-Hawking state of the two-sided AdS system in JT 
gravity. Green denotes an asymptotic boundary while  the violet one is the geodesic connects to two different points on the asymptotic boundary. {\it Right}: The wavefunction of the Hartle-Hawking state of a trumpet. The closed geodesic boundary is depicted in red.}
\label{fig:diskwavefunction}
\end{figure}

Now the aim is to use the Hartle-Hawking wavefunction 
construction \cite{Harlow:2018tqv} to explore certain features of the model and
in particular, to compute complexity. As we mentioned before, the fixed geodesic length between two parts of the asymptotic boundary by is denoted by $\ell$. 
Then the Hartle-Hawking wavefunction 
$\Phi_{\pm}(\beta,\ell)$ corresponding  
 to the integral over all Euclidean geometries with disk topology and asymptotic  boundary of renormalised length $\beta$ ( see left panel of fig.\ref{fig:diskwavefunction}) is given by
\cite{Fan:2021wsb}
\be\label{Dwf}
\Phi_{D,\pm}(\beta,y)=\frac{\sqrt{2}}{\pi}e^{S_0/2}\int_0^\infty dk e^{-\beta k^2}\cosh(2\pi k)\;\psi_{k,\pm}(z)
\,.
\ee
where 
\begin{align}
\label{eq:psikpm}
\psi_{k,\pm}(z)=\sqrt{z}\left(K_{2ik+\frac{1}{2}}(z)
\pm K_{2ik-\frac{1}{2}}(z)\right),
\end{align}
which follows from \eqref{eq:wavefunction-N1} through the identification $\ell =-\ln( z/2)$.
By gluing two such wavefunctions along the geodesic line $\ell$, one obtains the disk partition function as
\be\label{ZD}
Z_D(\beta)=\int_0^\infty\frac{dz}{z}\Phi_{D,\pm}^*(\beta/2,z)
\Phi_{D,\pm}(\beta/2,z)=e^{S_0}
\int_0^\infty dE \, e^{-\beta E}
\rho_0(E)=e^{S_0}\left(\frac{\pi}{4\beta}\right)^{1/2}e^{\frac{\pi^2}{\beta}}\,,
\ee
where the density of state is  
\be\label{den}
\rho_0(E)=\frac{\cosh 2\pi\sqrt{E}}{2\sqrt{E}}\,.
\ee
The wavefunction is normalised in such a way that it gives 
the correct expression for the partition function.

In order to consider the higher genus contributions, the most important 
ingredient is the wavefunction on the trumpet whose
asymptotic boundary is  pinched off by the disk region considered  in  left panel of fig.\ref{fig:diskwavefunction}. While more complicated hyperbolic surfaces require 
the use of Riemann surfaces with geodesic boundaries, the simplest configuration on the trumpet is depicted in right panel of fig.\ref{fig:diskwavefunction}.

The corresponding wavefunction $\Phi_{T,\pm}(\beta,b,z)$ can be realised as the trumpet partition function
pinched off by the disk wavefunction \eqref{Dwf}  through the
identity
\be
\int_0^\infty \frac{dz}{z}\Phi^*_{D,\pm}(u,z)\Phi_{T,\pm}(\beta-u,b,z)=Z_{T}(\beta,b)\,,
\ee
where the trumplet partition function is given by
\cite{ Stanford:2017thb, Fan:2021wsb}
\be
\label{eq:ZT-N1}
Z_{T}(\beta,b)=\int_0^\infty dk \cos( b k)\;e^{-\beta k^2}=\left(\frac{\pi}{4\beta}\right)^{1/2}e^{-\frac{b^2}{4\beta}}\,.
\ee
From this one arrives at
\be\label{Twf}
\Phi_{T,\pm}(\beta,b,z)=\frac{\sqrt{2}}{\pi}e^{-S_0/2}\int_0^\infty dk \cos( bk)\; e^{-\beta k^2}\psi_{k,\pm}(z)\,.
\ee
Once we have the disk and the trumpet wavefunctions, it is straightforward to compute disk and trumpet propagators which are defined respectively as follows
\begin{align}
\label{eq:ZDT-N1}
Z_{D}(\beta)&=\int \frac{dz_1}{z_1}\;\frac{dz_2}{z_2}
\Phi_{D,\pm}^*(\beta_1,z_1)P_{D,\pm}(u_1,u_2,z_1,z_2)
\Phi_{D,\pm}^*(\beta_2,z_2), \nonumber \\
Z_{T}(\beta,b)&=\int \frac{dz_1}{z_1}\;\frac{dz_2}{z_2}
\Phi_{T,\pm}^*(\beta_1,z_1)P_{T,\pm}(u_1,u_2,b,z_1,z_2)
\Phi_{T,\pm}^*(\beta_2,z_2),
\end{align}
with the condition $\beta=\beta_1+\beta_2+u_1+u_2$.

Using \eqref{ZD} and \eqref{eq:ZT-N1} in \eqref{eq:ZDT-N1}, one obtains the disk and the trumpet propagators as
\begin{align}
&&P_{D,\pm}(u_1,u_2,z_1,z_2)=\int_0^\infty dk e^{-(u_1+u_2)k^2}\cosh(2\pi k)\;
\psi_{k,\pm}(z_1)\;\psi_{k,\pm}(z_2),\cr &&\cr
&&P_{T,\pm}(u_1,u_2,b,z_1,z_2)=\int_0^\infty dk e^{-(u_1+u_2)k^2}\cos(b k)\;
\psi_{k,\pm}(z_1)\;\psi_{k,\pm}(z_2),
\end{align}

\subsection{Correlation functions}

Using the wavefunctions  presented in the previous subsection, we will now study correlation functions in the ${\cal N} = 1$ supersymmetric JT gravity model. We will first compute the spectral form factor and then will extend our study to the matter correlation functions as well.

The starting point is to evaluate the partition function $Z(\beta)$ which, at the 
leading order, is given by disk partition function and gets corrected further by higher genus contributions. Schematically, one can organize the full partition function as
\be
Z(\beta)=Z_D(\beta)+Z_f(\beta)+Z_1(\beta)+\cdots\,,
\ee
where the disk partition function, $Z_D(\beta)$ is given by \eqref{ZD} and $Z_f(\beta)$ is a contribution to the partition function coming from an external defect, parametrized by a function $f(b)$. The latter can be obtained from the trumpet partition function whose 
closed geodesic $b$ is capped off by a function $f(b)$
\be
Z_f(\beta)=\int_0^\infty db\; f(b) Z_T(\beta, b)=\int_0^\infty dE\;e^{-\beta E} \rho_f(E)\,,\;\;\;\;\;\;\rho_f(E)=\int_0^\infty db \frac{f(b)\cos( b \sqrt{E})}{2\sqrt{E}}\,.
\ee 

Of course, we can have different functions $f(b)$  which correspond to 
different physical objects we may wish to add to our model. In particular, taking $f(b)=\pm e^{-b\zeta}$ gives the contribution of a (anti)brane with tension $\zeta$ \cite{Okuyama:2021eju},
while the choice $f(b)=\frac{1}{\tanh{ \frac{b}{4}}}$ corresponds to the crosscap contribution
 \cite{Yan:2022nod}. Having known the explicit form of $Z_T(\beta,b)$, it is relatively simple to perform the integration over $b$, {\it e.g.} for $f(b) = e^{-\zeta b}$ one obtains
 \be
 \label{eq:partition-exp-N1}
 Z_f(\beta)=\frac{\pi}{2}e^{\beta\zeta^2}{\rm Erf}(\zeta \sqrt{\beta})\,.
 \ee

$Z_1(\beta)$ denotes the one loop contribution obtained by gluing two trumpets as
\be
\label{Z1}
Z_1(\beta)=\int \frac{dz}{z}\int b db\; \Phi^*_{T,\pm}(u,b,z)\Phi_{T,\pm}(\beta-u,b,z)=e^{-S_0}\int_0^\infty dE\;e^{-\beta E} \rho_1(E),
\ee
where
\be
\rho_1(E)= \frac{1}{\rho_0({E})} \int_0^\infty bdb\frac{\cos^2( b\sqrt{E})}{4E},
\ee
which is  divergent and needs to be regularized. We will be back to this point later in this section. The generalization for higher loops is also evident. The higher order contributions are denoted by the dots on the r.h.s of \eqref{eq:partition-exp-N1}.

Another quantity of interest is the two-point function, 
\begin{align}\label{eq:ZZ}
\langle Z(\beta_1)Z(\beta_2)\rangle=& \int_0^\infty b_1db_1\;b_2db_2\;
Z_T(\beta_1,b_1)X(b_1,b_2)Z_T(\beta_2,b_2)\\
=&\int_0^\infty dk_1dk_2\;
e^{-\beta_1 k_1^2-\beta_2 k_2^2}
\int_0^\infty b_1db_1 b_2db_2 X(b_1,b_2) 
{\cos(k_1b_1)\cos(k_2b_2)}.\nonumber
\end{align}
Here the function  $X(b_1,b_2)$ denotes the topologically weighted sum 
over the Weil-Petersson volumes associated with surfaces having two geodesic boundaries. These boundaries are parametrized by $b_1$ and $b_2$. The two-point function is of the form
\be\label{eq:X}
X(b_1,b_2):= \sum_{g=0}e^{(2-2g) S_0} \left({ V}_{g-1,2}(b_1,b_2)+\sum_{a\geq 0}
{ V}_{g-a,1}(b_1)
{V}_{a,1}(b_2)\right)\,.
\ee
The first and the second terms of \eqref{eq:X} correspond to the connected and the 
disconnected contributions respectively. However, the moduli space volumes 
$V_{g=0,1}(b)$ and $V_{g=0,2}(b_1,b_2)$ appearing in \eqref{eq:X} are undefined. So we need to add the corresponding contributions in \eqref{eq:ZZ} by hand. For the disconnected contributions involving 
$V_{g=0,1}(b)$, one should use \eqref{ZD} 
while the two boundary connected contribution for $g=0$ should be introduced in the two-point function in the form
\begin{align}
     Z(\beta_1,\beta_2)_{g=0,n=2,\mu} &=\int_{0}^{\infty}b_1 db_1 b_2 db_2 
     Z_T(\beta_1,b_1) Z_T(\beta_2,b_2)\,.
\end{align}
In order to obtain the spectral form factor, one needs to analytically continue
\eqref{eq:ZZ} and express the resulting expression in terms of the energy variables. This procedure yields
\be\label{eq:SFF}
\langle Z(\beta+it)Z(\beta-it)\rangle=\int_0^\infty dE_1dE_2
e^{-\beta( E_1+E_2)-it(E_1- E_2)}\;
\langle \rho(E_1) \rho(E_2) \rangle
\ee
with
\be
\langle \rho(E_1)\rho(E_2)\rangle=
\int b_1db_1\,b_2db_2X(b_1,b_2)\;\frac{\cos(b_1\sqrt{E_1})\cos(b_2\sqrt{E_2})}{4\sqrt{E_1E_2}}
\ee
being the density of states corresponding to two boundary case. Taking into account 
the disconnected term one has 
\be
\langle \rho(E_1)\rho(E_2)\rangle=e^{2S_0}\rho_0(E_1)\rho_0(E_2)
+\langle \rho(E_1)\rho(E_2)\rangle_{\rm cont}.
\ee
At late times, the dominant contribution to the integral \eqref{eq:SFF} comes from the small energy ranges
$|E_1-E_2|\ll1$. In the bosonic JT gravity it was shown that, in the limit $|E_1-E_2|\ll1$,
the non-perturbative contributions yield the following expression for
two point function of the density of state \cite{Saad:2019lba}
\be \label{eq:sinekernel}
\langle \rho(E_1) \rho(E_2) \rangle \approx e^{2 S_0}\hat{\rho}_0(E_1)
\hat{\rho}_0(E_2)+e^{S_0}\hat{\rho}_0(E_2)\delta(E_1-E_2)
-\frac{\sin^2\left(\pi e^{S_0} \hat{\rho}_0(E_2)(E_1-E_2)\right)}{\pi^2 (E_1-E_2)^2}\,,
\ee
where ${\hat \rho}_0(E)$ corresponds to the genus zero bosonic density of states given by
 \be
 \hat\rho_0(E)=\frac{\sinh 2\pi\sqrt{2E}}{2\pi^2}\,.
 \ee
The last term in \eqref{eq:sinekernel} is the so-called sine-kernel. This term is crucial to get  a ramp-plateau structure for the spectral form factor. It turns out that the same term is also responsible to get a similar structure in the late time behaviour of complexity \cite{Iliesiu:2021ari, Alishahiha:2022kzc}. We will come back to the latter in the next section.

In order to repeat this computation of the spectral form factor for the ${\cal N}=1$ supersymmetric  case, we essentially need the expression for the two-boundary 
density of state, analogous to \eqref{eq:sinekernel}. Of course, to the best 
of our knowledge, such an expression has not been found before. However, following the result of one boundary density of state obtained for a particular ${\cal N}=1$ model in \cite{Stanford:2017thb} one can expect that in our ${\cal N}=1$ supersymmetric case as well, one could still get a non-perturbative term in the form of a sine-kernel. 

Therefore to proceed, up to local term,  we propose the following expression 
for two point function of density of state in $|E_1-E_2|\ll1$ limit
\be \label{eq:sinekernel-SUSY}
\langle \rho(E_1) \rho(E_2) \rangle \approx e^{2 S_0}{\rho}_0(E_1)
{\rho}_0(E_2)
-\frac{\sin^2\left(\pi e^{S_0} {\rho}_0(E_2)(E_1-E_2)\right)}{\pi^2 (E_1-E_2)^2}\,.
\ee
where $\rho_0(E)$ is density of state for the ${\cal N}=1$ supersymmetric case given in 
\eqref{den}. Actually to propose such an expression, we are encouraged from the
results we get from late times behavior of spectral form factor\footnote{More generally, such a form of the connected part of the spectral correlator can be argued from the fact that at late times, the leading connected contribution to the spectral correlation comes from the short range correlation which yields a universal sine-kernel given by \cite{Mehta}
\be
\label{eq:sine-kernel}
\rho_c(E_1,E_2) \sim -\frac{\sin^2\left(D (E_1 - E_2) \,\rho(E_1+E_2)\right)}{D^2\,(E_1-E_2)^2}
\ee
in the coincident limit $E_1 \rightarrow E_2$.
Here $D$ is the dimension of the Hilbert space, $D \sim e^{S_0}$.} (similar arguments will be elaborated in the context of complexity in section \ref{sec:comment-saturation}). As should be expected, plugging
\eqref{eq:sinekernel-SUSY} into \eqref{eq:SFF}, gives a ramp-plateau structure 
for the connected, and decaying behaviour for the disconnected contribution.

Using the wavefunction formalism, one can also compute the full quantum gravity 
expressions for the matter correlation functions coupled to matter fields. 
Let us recall that, in the bosonic case, we start from the correlation functions of operators near the boundary that are function of the boundary coordinate
$x_i$'s. There is also a scaling that is given by the holographic coordinate $\xi$.
In other words we start with typical correlators in the following form
\be
\prod_i \xi_i^{\Delta_i}\langle {\chi}_1(x_1)\cdots{\chi}_n(x_n) \rangle\,,
\ee
where $\Delta_i$ denotes the scaling dimension of the operator $\chi_i$.
Then this correlation function is dressed by the kernels that are functions of $u_i,\xi_i,x_i$
with $u_i$ being the emergent boundary time. In order to write the kernels, one needs to define new variables associated with those appearing in the wavefunction.
In the bosonic case we have $y$ and $\beta$ that should be mapped to
\be
\beta\rightarrow u,\;\;\;\;\;\;\;\;\;y^{-1}\rightarrow \frac{\omega}{4}= \frac{ x_1 - x_2   }{4\sqrt{\xi_1\xi_2}}
\ee
Following \cite{Yang:2018gdb}, at the disk level, the quantum gravity correlators take the form
\bea\label{eq:quantumgravitycorrelators}
\langle \chi_1(u_1)\cdots \chi_n(u_n)\rangle_D= \int_{x_1>\cdots>x_n}
\frac{\prod_{i}d\xi_idx_i}{{\rm Vol}\left(\text{SL}(2,R)\right)}\!&&\!K(u_{12},{\bf x}_1,{\bf x}_2)\cdots
K(u_{1n},{\bf x}_n,{\bf x}_1)\\ &&\!\times\,\prod_i \xi_i^{\Delta_i-2}\langle \chi_1(x_1)\cdots 
\chi_n(x_n)\rangle_{\rm CFT}\,.\nonumber
\eea
 
Here ${{\rm Vol}\left(\text{SL}(2,R)\right)}$ appears because of the unfixed $\text{SL}(2, R)$ 
gauge symmetry. The generalization of this correlation function for the ${\cal N} = 1$ has been done in \cite{Lin:2022zxd}. Using the results of \cite{Lin:2022zxd}, after gauge-fixing, and using the geodesic approximation for the holographic CFT correlator, namely,
\be
\label{eq:geo-approx}
\langle \chi_1(x_1)\chi_2(x_2)\rangle= e^{-\Delta \ell}\,,
\ee
two point function of an operator $\chi$ with scaling dimension $\Delta$  is given by
\be\label{bi-l}
\langle \chi(u)\chi(0)\rangle=\int\frac{dz}{z}\int_0^\infty b_1db_1\;b_2db_2\;
\Phi^*_{T,\pm}(u,b_1,z) X(b_1,b_2)\Phi_{T,\pm}(\beta-u,b_2,z) z^{2\Delta}.
\ee
Making use of \eqref{Twf} and \eqref{eq:psikpm} we arrive at
\bea
\langle \chi(u)\chi(0)\rangle =
\frac{2e^{-S_0}}{\pi^2}
\int_0^\infty dE_1\;dE_2 e^{-uE_1-(\beta-u)E_2}\langle \rho(E_1)\rho(E_2)\rangle
{\cal M}(\Delta,\sqrt{E_1},\sqrt{E_2}),\, \nonumber\\
\eea
where 
\be
\label{eq:matrixM}
{\cal M}(\Delta,\sqrt{E_1},\sqrt{E_2})=
\frac{|\Gamma(\frac{1}{2}+\Delta+i(\sqrt{E_1}-\sqrt{E_2}))\Gamma(\Delta+i(\sqrt{E_1}+\sqrt{E_2}))|^2+(\sqrt{E_2}\rightarrow -\sqrt{E_2})}{2^{2-2\Delta}\Gamma(2\Delta)}\,.
\ee
With the analytic continuation $u=\frac{\beta}{2}+it$ and using our proposed behavior for the two point function of the density of state at late times \eqref{eq:sinekernel-SUSY}, one obtains the behavior of the matter two point correlation function at late times. We find that the late time behaviour of the matter correlation function is qualitatively the same as the ramp-plateau structure of the spectral form factor. 

As an aside, we can also compute the crosscap contribution to the matter two point function. A disk with a crosscap is an example of non-orientable geometries. Topologically, this is equivalent to a Möbius strip. The importance of this configuration is that it results in a non-decaying part already in the disk level two-point correlator \cite{Yan:2022nod}. It can therefore be thought of as one of the candidates to explain the original information puzzle by Maldacena \cite{Maldacena:2001kr} posed in terms of the conflict of the exponential decay of holographic correlators with the unitarity of the boundary CFT. Indeed, following \cite{Yan:2022nod}, the corresponding contribution is given by
\be
\langle \chi(u)\chi(0)\rangle_{\rm crosscap}=\int_0^\infty \frac{dz}{z}
P_{T,\pm}(u,\beta-u,z,z) \, z^{2\Delta}=\int_0^\infty dE\,e^{-\beta E} \rho_0(E)
{\cal M}(\Delta,\sqrt{E},\sqrt{E})\,,
\ee
which is constant, {\it i.e.} independent of $u$ and is of ${\cal O}(1)$.

While deriving the matter correlation function, we used the geodesic approximation \eqref{eq:geo-approx} and therefore, strictly speaking, this computation is valid only for large $\Delta$. However, one can also note that setting $\Delta=0$ in the expression for
two point function \eqref{bi-l} reduces to that of the partition function \eqref{Z1}.
From this observation, one can find a procedure to regularize the partition function. Of course, the main subtlety that still remains is to find full and exact expression for two point function of the density of state. 
\subsection{Complexity via replica trick}
In this section, we will study the late time behaviour of complexity in our setup. 
As discussed in the introduction, in two dimensions, the holographic complexity via the CV proposal \cite{Stanford:2014jda} amounts to computing the length of a geodesic connecting two points on the asymptotic AdS boundary. 
Denoting the classical geodesic distance denoted by $\ell=- \ln z/2$, it is natural to expect complexity to be proportional to the quantum expectation value of the geodesic ${\cal C}\sim \langle \ell\rangle_{\rm QG}$ in a theory of quantum gravity.

This definition was used to compute the late time 
behaviour of complexity of a one-sided 
black hole in \cite{Alishahiha:2022kzc}. It was shown that the complexity 
exhibits linear growth at late times before eventually saturating to a finite value. 
The key step in this construction was the use of the non-perturbative expression for the two-point function of the density of state to obtain the saturation at late times. Although complexity of a two-sided has also been studied in \cite{Iliesiu:2021ari}, for the reasons explained in the introduction, we will use the formalism developed in \cite{Alishahiha:2022kzc} to compute the late-time behaviour of complexity for the supersymmetric models we consider in this work. 

To proceed, follwoing  \cite{Iliesiu:2021ari,Alishahiha:2022kzc}, we note that in order to see the saturation phase, one needs  to compute the quantum expectation of geodesic length taking into account surfaces of higher genus. By making use of the trumpet wavefunctions \eqref{Twf}, one has\footnote{In what follows we will study the time dependence of this 
quantity for  the ``plus'' sign
wavefunction we obtained in the previous subsection and therefore we will drop
the ``$\pm$'' from now on.
}
\be
\langle \ell(u) \rangle =\frac{1}{Z(\beta)}\int d_1db_1\;b_2db_2 X(b_1,b_2)
\int_0^\infty\frac{dz}{z}\Phi_{ T}^*(u,b_1, z)\Phi_{ T}(
\beta-u,b_2,z)\left(-\ln\frac{z}{2}\right)\,.
\ee
To compute the complexity we use the replica trick \eqref{RT} to express the quantum length as
\be
\label{RT2}
\langle \ell(u) \rangle =-\lim_{N\rightarrow 0}\frac{\langle z^{2N}\rangle-1}{N}
=-\lim_{N\rightarrow 0}\partial_N \langle z^{2N}\rangle
\ee
where
\be
\langle z^{2N} \rangle =\frac{1}{2Z(\beta)}\int d_1db_1\;b_2db_2 X(b_1,b_2)
\int_0^\infty\frac{dz}{z}\Phi_{T}^*(u,b_1, z)\Phi_{ T}(
\beta-u,b_2,z)\left(\frac{z}{2}\right)^{2N}\,.
\ee
Making use of the  explicit form of the trumpet wavefunction, \eqref{Twf} one 
arrives at

\bea
\langle z^{2N} \rangle_u =
\frac{e^{-S_0}}{\pi^2Z(\beta)}
\int_0^\infty dE_1\;dE_2 e^{-uE_1-(\beta-u)E_2}\langle \rho(E_1)\rho(E_2)\rangle
{\cal M}(N,\sqrt{E_1},\sqrt{E_2})\,
\eea
which has the same form that that of matter two-point function \eqref{bi-l}
once we identify $N$ with the dimension of an operator, $\Delta$. Of course, 
we should not consider this as the two-point function since the $N\rightarrow
0$ limit we consider in \eqref{RT2} is the opposite to the limit of geodesic approximation, namely the large $\Delta$ limit which yields the semi-classical description of the matter two-point function \eqref{bi-l}.

In order to evaluate the late time behaviour of complexity, the scheme is as follows.
First we need to make the analytic continuation $u=\frac{\beta}{2}+it$ and
then plug the
resulting expression in the replica formula and take the $N\rightarrow 0$ limit. Furthermore since we are interested in the late-time regime, we need to consider the coincident limit which is conveniently achieved by using a new set of variables $\{E, \omega \}$
defined as $E=\frac{E_1+E_2}{2},
\omega=E_1-E_2$. In these coordinates, one finds
\be
\langle z^{2N} \rangle_t =
\frac{e^{-S_0}}{\pi^2Z(\beta)}
\int_0^\infty dE\int_{-\infty}^\infty d\omega e^{-\beta E-it\omega}\langle \rho(E+\omega/2)\rho(E-
\omega/2)\rangle
{\cal M}(N,\sqrt{E+\omega/2},\sqrt{E-\omega/2})\,.
\ee
Now at late Lorentzian times, clearly, the main contribution comes from the coincident limit {\it i.e} $\omega\rightarrow 0$. In this limit we have
\be
\lim_{N\rightarrow 0}\partial_N {\cal M}(N,\sqrt{E+\omega/2},\sqrt{E-\omega/2})
\approx \frac{\pi\sqrt{E}}{\rho_0(E)}\,\frac{1}{\omega^2}+{\rm local\, terms}\,.
\ee
Plugging this back to the replica formula \eqref{RT2}, one arrives at
\be
\label{eq:lt}
\langle \ell(t)\rangle ={\rm Constant}-
\frac{e^{S_0}}{\pi Z(\beta)}
\int_0^\infty dE e^{-\beta E} {\sqrt{E}}\rho_0(E)\int_{-\infty}^\infty d\omega 
\frac{e^{-it\omega}}{\omega^2}\left(1-\frac{\sin^2\left(\pi e^{S_0} {\rho}_0(E)\omega\right)}{(\pi\rho_0(E)e^S \omega)^2}\right)
\,.
\ee
We note that this is very similar to the bosonic case apart from the fact that 
in this expression, the density of state $\rho_0$ is given by that of supersymmetric case
\eqref{den}. Therefore we can conclude that at late
times when $\omega\sim \frac{1}{t}\rightarrow 0$ and 
for $\rho_0 \, \omega\gg 1$ essentially the first term in the bracket on the r.h.s of  \eqref{eq:lt} dominates leading 
to a linear growth, while for $\rho_0 \, \omega\ll  1$ which is the case at $t \sim e^{S_0}$, the second term starts dominating that essentially cause the whole integral to approach zero
leading to a constant complexity. More precisely for $\pi \rho_0(E)e^{S_0}\omega\gg
1$, expression the sin-term in terms of the exponentials one 
finds \cite{Iliesiu:2021ari,Alishahiha:2022kzc}
\be
\int_{-\infty}^\infty d\omega 
\frac{e^{-it\omega}}{\omega^2}\left(1-\frac{\sin^2\left(\pi e^{S_0} {\rho}_0(E)\omega\right)}{(\pi\rho_0(E)e^S \omega)^2}\right)=\frac{2\pi^2\rho_0(E)e^{S_0}}
{3}\left(1-\frac{t}{2\pi\rho_0(E)e^{S_0}}\right)^3,
\ee
which can be used to recast \eqref{eq:lt} into the following form
\be
\label{eq:lt1}
\langle \ell(t)\rangle ={\rm Constant}-
\frac{2\pi e^{2S_0}}{3Z(\beta)}
\int_{E_0}^\infty dE e^{-\beta E} {\sqrt{E}}\rho^2_0(E)\left(1-\frac{t}{2\pi\rho_0(E)e^{S_0}}\right)^3
\,,
\ee
where $E_0$ is a solution of the equation $\pi \rho_0(E)e^{S_0}
\omega=1$.
It is now straightforward to perform $E$-integral. Indeed for  $t\ll e^{S_0}$ 
expanding the bracket one gets the linear growth.  On the other hand for 
times of order of $t\sim e^{S_0}$ the lower limit of the above integral tends to 
infinity that results in the vanishing of the integral, demonstrating  the saturation phase.

In this way, just like the bosonic case, we obtain a late time saturation of complexity for the ${\cal N} = 1$ supersymmetric model as well, due to an explicit cancellation between the disk level linear growth and the non-perturbative contribution.

So far we have studied the late- time behaviour of complexity and have shown that it 
exhibits a saturation phase at late times. We note, however, that an important 
feature of complexity is that there are complexity fluctuations around the saturation value that are of ${\cal O}(1)$. Therefore it is important to compute the variance of the complexity above the saturation value\cite{Iliesiu:2021ari,Alishahiha:2022kzc}.

Following \cite{Alishahiha:2022kzc}, the relevant part contributing to the variance can be  
obtained from quantum expectation value of $\langle \ell^2(u)\rangle_{\rm C}$ that is given by
\bea
&&\langle \ell^2(u) \rangle_{\rm C} =\frac{1}{Z(\beta)}\int d_1db_1\;b_2db_2 X(b_1,b_2)
\int_0^\infty\frac{dz}{z}\Phi_{ T}^*(u,b_1, z)\Phi_{ T}(
\beta-u,b_2,z)\ln^2\frac{z}{2}\\
&&\;\;\;\;\;\;\;\;\;\;\;\;=\frac{1}{4Z(\beta)}\lim_{N\rightarrow 0}\partial^2_N\int d_1db_1\;b_2db_2 X(b_1,b_2)
\int_0^\infty\frac{dz}{z}\Phi_{ T}^*(u,b_1, z)\Phi_{ T}(
\beta-u,b_2,z)\left(\frac{z}{2}\right)^{2N},\nonumber
\eea
that is  
\bea
\langle \ell^2(u) \rangle_{\rm C} =
\frac{e^{-S_0}}{4\pi^2Z(\beta)}
\int_0^\infty dE_1\;dE_2 e^{-uE_1-(\beta-u)E_2}\langle \rho(E_1)\rho(E_2)\rangle
\lim_{N\rightarrow 0}\partial^2_N{\cal M}(N,\sqrt{E_1},\sqrt{E_2})\, \nonumber\\
\eea
Here the subscript \text{C} denotes the connected contribution, namely,
\be\label{eq:variancetwosided}
\langle\ell^2(u)\rangle_{\text{C}} = \langle\ell^2(u)\rangle-
\langle\ell(u)\rangle^2\,.
\ee

Using explicit form of ${\cal M}$ given in \eqref{eq:matrixM}, one can compute the variance to see that although 
it is not of order of one, it remains constant at the late times and eventually saturates at a value which is of order of the square root of the complexity. 
Indeed, the result is qualitatively the same
as that obtained in the case of bosonic JT gravity\cite{Alishahiha:2022kzc}.

Before ending this section, it is worth noting that the quantity given in the equation \eqref{bi-l}, is the higher genus contribution to
the correlation function of the bosonic part of a bi-local operator whose disk 
level correlation has been studied in
\cite{Mertens:2017mtv,Fan:2021wsb}. In our notation, the disk level is actually the disconnected part of the expression \eqref{bi-l} in which correlation function of two density of state should be replaced by factorized product $e^{S_0}\rho_0(E_1) \rho_0(E_2)$. 

We should nevertheless remember that the corresponding 
operator also has a fermionic counterpart whose correlation function at disk level is given by
\be
\langle {\cal O}_f(u) \rangle =\frac{e^{-S_0}}{\pi^2 Z_D(\beta)}
\int_0^\infty dk_1\;dk_2 e^{-uk_1^2-(\beta-u)k_2^2} \cosh(2\pi k_1)\cosh(2\pi k_2)
{\hat {\cal M}}(N,k_1,k_2)\,,
\ee
where 
\be
{\hat{\cal M}}(N,k_1,k_2)=
\frac{(k_1+k_2)^2|\Gamma(\frac{1}{2}+N+i(k_1-k_2))\Gamma(N+i(k_1+k_2))|^2+(k_2\rightarrow -k_2)}{4\Gamma(2N)}\,.
\ee
We note that due to the prefactor $(k_1-k_2)^2$ in front of the Gamma functions,
unlike the bosonic part, the derivative of this quantity with respect to $N$ does not 
exhibit linear growth at the disk level. The situation is very similar to that of
correlation function itself. 

Of course, a priori there is no good reason why one should expect the replica trick to be applicable for the fermionic part of the bi-local operator. Nonetheless, taking into account the higher genus contribution one can still get a
late-time linear growth followed by a saturation phase. Therefore, the physics behind
this saturation is more similar to that of spectral form factor or entanglement, than that of the complexity. It is worth exploring this curious fact better in future.


\section{JT gravity with ${\cal N} = 2$ supersymmetry}
\label{sec:N2}
In this section we will extend our analysis the another supersymmetric model with more supercharges. For the ${\cal N}=1$ case we dealt with a pair of Majorana fermions acting
either on the left or on the right side, which were understood as the supersymmetric partners of the bosonic variable given by the normalized length $\ell$.
For the ${\cal N}=2$ supersymmetric JT gravity model which we are going to discuss in this section, one has two supersymmetries acting on both sides. Additionally, in this
case one also has a compact scalar field. 

In the ${\cal N}=2$ model we consider here, the Hamiltonian describing the system is given by \cite{Lin:2022zxd}\footnote{Our geodesic length $\ell$ is related to that in \cite{Lin:2022zxd} by a constant shift.}
\be \label{SH}
H = - \partial_\ell^2 +4 e^{-  \ell} - { \frac{1}{4} } \partial_a^2 + 2 i \left(  \bar \psi_l  \psi_re^{- ia} +  \bar \psi_r   \psi_l   e^{ i a}  \right) e^{-\ell/2}\,,
\ee 
where $\psi_{r,l}$  are the supersymmetric partners of $\ell$ and $a$ is a compact scalar corresponding to the phase associated to a relative $U(1)_R$ symmetry. Note that
due to the quantization of the $R$ symmetry, the scalar $a$ is periodic $a\sim a+
2\pi q$ with $1/q$ being the unit quantization of the $R$ charges.

This Hamiltonian \eqref{SH} can be obtained from the supersymmetry algebra as follows
\be 
\{ Q_r , Q_l \} = \{ \bar Q_r , \bar Q_l \}=0\,\;\;\;\;\;\;\;\;\;\;\{ Q_r , \bar Q_r \}= 
\{ Q_l, \bar Q_l\} =H\,, 
\ee 
where the supercharges are given by\cite{Lin:2022zxd}
\bea
Q_r &=&  \psi_r ( i \partial_\ell + \half \partial_a) + 2e^{ - \ell/2  + i a} \psi_l\,,\;\;\;\;\;\;\;\;\;\;\bar Q_r =  \bar \psi_r ( i \partial_\ell - \half \partial_a ) +  2e^{ -\ell/2  - i a} \bar \psi_l\,,
\cr 
Q_l &=&  \psi_l ( i \partial_\ell - \half \partial_a ) -2 e^{ - \ell/2 - i a} \psi_r\,,\;\;\;\;\;\;\;\;\;\; \bar Q_l =   \bar \psi_l ( i \partial_\ell +   \half \partial_a ) - 2 e^{- \ell/2 + i a  } \bar \psi_r.
\eea 
The $R$ symmetry generators acting on right and left sides are  
 \be \label{R-symmetry}
J_r =- i \partial_a - \half [ \bar \psi_r, \psi_r]\,,\;\;\;\;\;\;\;\;\;\;J_l =  i  \partial_a - \half [ \bar \psi_l, \psi_l]\,.
\ee 
 The total fermion number can be expressed in terms of the $R$ symmetry generators as
\be
(-1)^{F}=e^{q\pi i J_r }\,e^{q\pi i J_l }\,.
\ee

Now the aim is to solve the Schr\"odinger equation associated to the supersymmetric Hamiltonian \eqref{SH}. Being supersymmetric, one needs to take care of the fermionic
structure of the equation. To do so, we start with a fermionic vacuum solution 
that is killed by left and right fermionic operators $\psi_{l,r}$ \cite{Lin:2022zxd}

\be \label{Vf}
\psi_{r,l}\ \left|\half,\half \right\rangle = 0\,. 
\ee 
It is a state with $R$ charges $J_l = J_r = \half$. It has odd fermion numbers as
$J_l + J_r=1$. Since we have two supercharges, for any state with positive energy we
get a multiplet with four members, two of which have odd fermionic number.
Using the above fermionic vacuum, it is easy to construct all members of the multiplet 
by acting with $\bar Q_{l,r}$. 

To find the full wave function including the  bosonic part, we consider the following ansatz
\be
|\psi\rangle=e^{2ima}f(\ell) \left|\half,\half \right\rangle\,.
\ee
Here we have used the fact that the scalar $a$ is a periodic function. One can note that this state is killed by $Q_{l,r}$ and its $R$ charges are $j_r=m+\frac{1}{2}$ and
$j_l=-m+\frac{1}{2}$, thus having odd fermion number.

Plugging this ansatz in the Schr\"odinger equation associated with the Hamiltonian
\eqref{SH}, one arrives at
\be
(-\partial_\ell+4e^{-\ell})f(\ell)=(E- m^2)f(\ell)\,,
\ee
solutions whereof are given by 
\be
f_{k}(y)=K_{2ik}(y),\;\;\;\;\;\;\;{\rm with}\;\;\; k^2=E- m^2\,,
\ee
where we have set $y=2e^{-\ell/2}$.

Therefore the highest  component of the desired supersymmetric multiplet with $j_l+j_r=1$ can be written as \footnote{We are using a notation in which states are labeled by
 $|j_r,j_l\rangle$.}
\be\label{HCom}
\left|m+\frac{1}{2},-m+\frac{1}{2}\right\rangle_k=e^{2ima}\sqrt{2r(k)}K_{2ik}(y) 
\left|\half,\half \right\rangle\,.
\ee 
Here $r(k)=\frac{8k\sinh 2\pi k}{\pi^2}$.

Using the supercharges $\bar Q_{r,l}$ one can obtain other elements of the multiplet as follows
\bea
&&\left|m+\frac{1}{2},-m-\frac{1}{2}\right\rangle_k=\frac{\bar Q_l}{\sqrt{E}}\left|m+\frac{1}{2},-m+\frac{1}{2}\right\rangle_k=e^{2ima}
\sqrt{\frac{r(k)}{2E}}f^l_k(y) \left|\half,\half \right\rangle\,,\\
&&\left|m-\frac{1}{2},-m+\frac{1}{2}\right\rangle_k=\frac{i\bar Q_r}{\sqrt{E}}\left|m+\frac{1}{2},-m+\frac{1}{2}\right\rangle_k=e^{2ima}\sqrt{\frac{r(k)}{2E}}f^r_k(y) \left|\half,\half \right\rangle\,,\cr &&\cr&&
\left|m-\frac{1}{2},-m-\frac{1}{2}\right\rangle_k=\frac{\bar Q_l\bar Q_r}{2E}\left|m+\frac{1}{2},-m+\frac{1}{2}\right\rangle_k=e^{2ima}\sqrt{2r(k)}f^{lr}_k(y) \left|\half,\half \right\rangle,\nonumber
\eea
with $f^{lr}_k(y)=K_{2ik}(y) \bar{ \psi}_l \bar{ \psi}_r$ and 
\bea
&&f^l_k(y)=i\bigg(yK_{2ik-1}(y)+2(ik+m)K_{2ik}(y)\bigg)\bar \psi_l-yK_{2ik}(y)e^{ia}\bar \psi_r,\cr &&\cr &&
f^r_k(\ell)=iyK_{2ik}(y)e^{-ia}\bar \psi_l-\bigg(yK_{2ik-1}(y)+2(ik-m)K_{2ik}(y)\bigg)\bar \psi_r
\,.
\eea
Note that, in our notation, all states are normalized\footnote{
Note that $\langle f^{l,r}_{k'}(y)|f^{l,r}_k(y)\rangle=\frac{\pi^2 E}{2k\sinh 2\pi k}\delta(k-k')$. This is used to fix the function r(k) in \eqref{HCom}.}
\be
\langle j'_l,j'_r|j_r,j_l\rangle=\delta_{mm'}\delta(k-k').
\ee
The state with zero energy may be obtained from the above states by setting $k=\pm im$. Due to symmetric properties of Bessel functions, one is free to choose any one  of the signs. In particular, setting $k=im$,  the zero energy state created by $\bar Q_l$ is given by
\be
\left|m+\frac{1}{2},-m-\frac{1}{2}\right\rangle_{im}=e^{2ima}\sqrt{\frac{4\sin
2\pi m}{\pi}}
f^l_{m}(y) \left|\half,\half \right\rangle\,,\;\;\;\;\;\;\;\;-\frac{1}{2}\leq m\leq 0
\ee
On the other hand, the zero energy state created by $\bar Q_r$ has the form
\begin{align}
\left|m-\frac{1}{2},-m+\frac{1}{2}\right\rangle_{im} =e^{2ima}\sqrt{\frac{4\sin
2\pi m}{\pi}}
f^r_{m}(y) \left|\half,\half \right\rangle\,,\;\;\;\;\;\;\;\;\;0\leq m\leq\frac{1}{2}
\end{align}
 Here 
\begin{align}
f^l_{m}(y) &= iyK_{2m+1}(y)\bar \psi_l-yK_{2m}(y)e^{ia}\bar \psi_r\,\cr\nonumber\\
f^r_{m}(y) &= iyK_{2m}(y)e^{-ia}\bar \psi_l-yK_{2m-1}(y)\bar \psi_r,.
\end{align}

\subsection{The Hartle-Hawking wavefunctions}
We have now all the ingredients to write done the Hartle-Hawking wavefuction on a disk for the ${\cal N} = 2$ supersymmetric model.
First we will compute the Hartle-Hawking wave function corresponding to the elements of the above multiplet with $j_l+j_r=0$. 
It is also very important to note that in the supersymmetric case one has a vacuum degeneracy when $E=0$. Therefore, while writing down the wavefunction, the zero energy contributions need to be taken into account separately. 

To write down the wavefunction it is useful to utilize Grassmannian variables to represent the spin structure and the inner products thereof in terms of Grassmann (also called the Berezin) integrals. Let us consider the Grassman variable $\chi$ and make the following identifications
\be
\bar \psi_r\rightarrow -\chi_1,\;\;\;\;\;\;\;\;\;\;\bar \psi_l\rightarrow -i\chi_2
\ee
with the inner product condition $\int d\chi_1\,d\chi_2 \;\chi_1\chi_2=1$. Moreover, 
we assume $\bar \chi_1=-\chi_2$ and $\bar \chi_2=\chi_1$.
In this notation, for a fixed value of $m$, the ${\cal N} = 2$ supersymmetric version of the Hartle-Hawking wavefunctions at the disk level, are given, for the left and the right sides as
\be
\label{eq:phi-mode-exp}
\Phi^{l,r}_m(\beta,y,a)=e^{S_0/2}e^{2ima}\left(-\frac{2}{\sqrt{\pi}}
\sin( 2\pi m) \hat f^{l,r}_{m}(y,a)+\int_0^\infty dk e^{-\beta E}
\frac{2k \sinh 2\pi k}{\pi E}\hat f^{l,r}_k(y,a)\right),
\ee
where $E=k^2+m^2$ and
\bea
\label{eq:f-lr}
&& \hat f^l_k(y,a)= \bigg(yK_{2ik-1}(y)+2(ik+m)K_{2ik}(y)\bigg) \chi_2+yK_{2ik}(y)e^{ia} \chi_1,\cr &&\cr &&
 \hat f^r_k(y,a)=\bigg(yK_{2ik-1}(y)+2(ik-m)K_{2ik}(y)\bigg)\chi_1+yK_{2ik}(y)e^{-ia}\chi_2\,,\cr &&\cr 
&&
\hat f^l_{m}(y,a)= yK_{2m+1}(y)\chi_2+yK_{2m}(y)e^{ia}\chi_1,\cr &&\cr 
&&
\hat f^{r}_{m}(y,a)= yK_{2m}(y)e^{-ia} \chi_2+yK_{2m-1}(y) \chi_1
\,.
\eea
The above wavefunction is a function of the geodesic length $y$ along which different 
wavefunctions can be glued through an integration over this length, with an appropriate
measure of integration. In our convention, the measure is $\int_0^\infty \frac{dy}{y}$. Of course, one also needs to perform an integration over $a$ that, as we will elaborate below, typically results in conservation laws appearing in the form the Kronecker $\delta$-functions. The measure 
of this integral is $\int_0^{2\pi q}\frac{da}{2\pi q}$ with $q$ being the period of $a$. In particular, one can compute the corresponding partition function of sector $m$
as 
\be
\label{eq:partion-disk1}
Z^{l,r}_{m,m'}(\beta)=\int_0^\infty \frac{dy}{y}\int_0^{2\pi q}\frac{da}{2\pi q}\int
d\chi_1 d\chi_2\;
\bar \Phi^{l,r}_{m'}(\beta/2,y,a)\Phi^{l,r}_m(\beta/2,y,a)
\ee
The strategy to perform this integration is as follows. We first integrate over the Grassmann variables $\chi_1$ and $\chi_2$ to get rid of the fermionic structure of the wavefunctions. We are then be left with non-trivial $a$-integrals that, due
to simplicity of  the $a$-dependence in the remaining terms, typically result in a delta function in the form of $\frac{1}{2}\delta_{mm'}$. Finally, using the orthogonality condition of the Bessel functions, we perform the integral over $y$ to get
\be
\label{eq:partition}
Z^{l,r}_{m}(\beta)=e^{S_0}\left(-\sin( 2\pi m)
 +
\int_0^\infty dk e^{-\beta E}\;
\frac{k\sinh 2\pi k}{E}\right)\,.
 \ee
 
Of course, finally one needs to sum over all $m$-sectors as well. However, as mentioned above, in this summation one needs also to include, separately, the zero energy part which should be taken for $-\frac{1}{2}\leq m\leq 0$ for the left side and $0\leq m\leq\frac{1}{2}$ for the right side.

It is important to note that the wavefunction associated with partition function \eqref{eq:partition} does not correspond to the wavefunction for the TFD state of a two sided black hole. In order to obtain the latter one needs to sum over both the left and the right sectors \cite{Lin:2022zxd}.
\be
\label{eq:TFD1}
\Psi_D^{\rm TFD}(\beta,\ell,a)=e^{S_0/2}\sum_m e^{2ima}
\hat f^{\rm TFD}_m(y,a),
 \ee
 where
 \be
 \label{eq:TFD2}
\hat f^{\rm TFD}_m(y)=-\frac{2}{\sqrt{\pi}}
\sin( 2\pi m)\hat f^{l}_{m}(y,a)+\int_0^\infty dk e^{-\beta E}
\frac{2k \sinh 2\pi k}{\pi E}(\hat f^{l}_k(y,a)+\hat f^{r}_k(y,a))\,.
 \ee
 
It is easy to check that computing the partition function by using \eqref{eq:TFD2} yields the well-known TFD partition function. It is worth mentioning here that in the construction of the TFD wavefunction, it is sufficient to consider the zero mode coming from only one side \cite{Lin:2022zxd}. This is because, these zero modes are secretly shared by both the right and the left sectors and are responsible for the entanglement resulting in the maximally entangled TFD state. 
 
\subsection{Complexity via replica trick for ${\cal N}=2$ model}

Having written the explicit form of the wavefunction for the ${\cal N} = 2$ model, in this section, we will compute the  complexity for this model. To do so, as in the previous case,
we need to compute  $\langle \ell\rangle$. Since in the present case the states
additionally carry $R$ charges, in what follows, we will compute the {\it matrix element} of complexity labeled by the $R$ charges, with the understanding that at the end of the day we will have to sum over them. To be precise, let us compute the complexity matrix element $\langle \ell(u)\rangle_{mm'}$ as follows.
 \be
\langle \ell(u)\rangle_{mm'}=-\frac{2}{Z^{l,r}_m(\beta)}\int_0^\infty \frac{dy}{y}\int_0^{2\pi q}\frac{da}{2\pi q}\int
d\chi_1 d\chi_2\;
\bar \Phi^{l,r}_{m'}(u,y,a)\Phi^{l,r}_m(\beta-u,y,a)\ln\left(\frac{y}{4}\right)
\ee
Again, as chalked out before, in order to evaluate this we use the modified replica trick summarized in \eqref{RT} and \eqref{eq:replica-formula} which amounts to computing in this case
\be
\langle y^{2N}\rangle_{u,mm'}=\frac{1}{Z^{l,r}_m(\beta)}\int_0^\infty \frac{dy}{y}\int_0^{2\pi q}\frac{da}{2\pi q}\int
d\chi_1 d\chi_2\;
\bar \Phi^{l,r}_{m'}(u,y,a)\Phi^{l,r}_m(\beta-u,y,a)\left(\frac{y}{4}\right)^{2N}.
\ee
To be concrete, in what follows, we only consider the left side, while its extension for the other side is evident. Using the mode expansion given in 
\eqref{eq:phi-mode-exp} and \eqref{eq:f-lr}, we get
\bea
&&\langle y^{2N}\rangle_{u,mm'} =\frac{e^{S_0}}{Z^{l}_m(\beta)}\int_0^\infty \frac{dy}{y}y^{2N}\int_0^{2\pi q}\frac{da}{2\pi q}\int
d\chi_1 d\chi_2\;e^{2i(m-m')a} \\&&\hspace{1.5cm}\times 
\bigg\{\frac{4}{{\pi}}
\sin( 2\pi m')
\sin( 2\pi m)\bar {\hat f}^{l}_{m'}(y,a) \hat f^{l}_{m}(y,a) 
 \nonumber\\&&\hspace{2cm}
-\frac{2}{\sqrt{\pi}}
\sin( 2\pi m')\int_0^\infty dk_2 e^{-(\beta-u) E_2}
\frac{2k_2 \sinh 2\pi k_2}{\pi E_2} \;\bar {\hat f}^{l}_{m'}(y,a)\hat f^{l}_{k_2}(y,a)
 \nonumber\\&&\hspace{2cm}
-\frac{2}{\sqrt{\pi}}
\sin( 2\pi m)\int_0^\infty dk_1 e^{-u E_1}
\frac{2k_1 \sinh 2\pi k_1}{\pi E_1}\;\bar{\hat f}^{l}_{k_1}(y,a) \hat f^{l}_{m}(y,a)
 \nonumber\\&&\hspace{2cm}
+\int_0^\infty dk_1 dk_2 e^{-u E_1-(\beta-u) E_2}
\frac{2k_1 \sinh 2\pi k_1}{\pi E_1}
\frac{2k_2 \sinh 2\pi k_2}{\pi E_2}\bar{\hat f}^{l}_{k_1}(y,a)\hat f^{l}_{k_2}(y,a)\bigg\},\nonumber
\eea
which, upon using the explicit form of the wavefunction through \eqref{eq:f-lr} and the following the order of integration mentioned before, yields
\bea
\label{eq:replicayN-1}
&&\langle y^{2N}\rangle_{u,mm'}=
\frac{e^{S_0}\delta{mm'}}{Z^{l}_m(\beta)}\bigg\{\frac{4}{{\pi}}
\sin^2( 2\pi m) \, {\cal M}^{(1)}_m(N)
\\ 
&&\hspace*{3.5cm}-\frac{2}{\sqrt{\pi}}
\sin( 2\pi m)\int_{m^2}^\infty dE_1 \left(e^{-(\beta-u) E_1} + e^{-u E_1}\right)
{\hat \rho}_{0,m}(E_1)\, {\cal M}^{(2)}_m(N, E_1)
\cr \nonumber\\
&&\hspace*{3.5cm}+\int_{m^2}^\infty dE_1 dE_2\;e^{-u E_1-(\beta-u) E_2} \, {\hat \rho}_{0,m}(E_1){\hat \rho}_{0,m}(E_2) {\cal M}^{(3)}_m(N, E_1, E_2) \bigg\},
\nonumber
\eea
where
\bea
&&{\cal M}^{(1)}_m(N) = 2^{2N - 3} \frac{N\Gamma^2(N)}{\Gamma(2N)}\Gamma(N-2m)\Gamma(N+1+2m)  \\
&&{\cal M}^{(2)}_m(N, E_1) = 2^{2N-2} (E_1+N(N+2m)) \frac{|\Gamma(N+m+ik_1)\Gamma(N-m+ik_1)|^2}{\pi \, \Gamma(2N)}\nonumber \\
&&{\cal M}^{(3)}_m(N, E_1, E_2) =  2^{2N-1}\,(E_1+E_2+N(N+2m))
\frac{|\Gamma(N+i(k_1+k_2)) \Gamma(N+i(k_1-k_2))|^2}{\pi^2 \,\Gamma(2N)},\nonumber
\eea
and ${\hat \rho}_{0,m}(E_i)$ is the effective density of states for a discrete $m$ mode, which follows directly from \eqref{eq:partition}.
\be
{\hat \rho}_{0,m}(E_i) = \frac{\sinh{2\pi\sqrt{E_i-m^2}}}{2 E_i},
\ee

Now we are in a position to evaluate the late time behaviour of complexity. For this, we need to follow the same procedure as adopted for the ${\cal N} = 1$ case above. Namely,
we need to first make the analytic continuation $u=\frac{\beta}{2}+it$. and then plug the resulting expression in the replica formula before finally taking the $N\rightarrow 0$ limit.
Furthermore, since we are interested particularly in the late time behaviour, we need to consider the coincident limit, namely $E_1 \rightarrow E_2$. This limit is implemented by using a new set of variables, $E=\frac{E_1+E_2}{2},
\omega=E_1-E_2$, and taking $\omega \rightarrow 0$ limit. In this limit one has
\bea
\label{eq:M-limits}
&&\lim_{N\rightarrow 0}\partial_N {\cal M}^{(1)}_m(N)
\approx -\frac{\pi}{4 \sin{2 \pi m}} \left(\psi ^{(0)}(-2 m)+\psi ^{(0)}(2 m+1)+\log (4)\right)\nonumber \\ &&
\lim_{N\rightarrow 0}\partial_N {\cal M}^{(2)}_m(N, E+ \omega/2)
\approx \frac{\pi}{\cosh{2\pi \sqrt{E - m^2}-\cos{2 \pi m}}}+{\rm local\, terms}\,.\nonumber \\ &&
\lim_{N\rightarrow 0}\partial_N {\cal M}^{(3)}_m(N,E+\omega/2,E-\omega/2)
\approx \frac{1}{{\hat \rho}_{0,m}(E)}\,\frac{2\sqrt{E-m^2}}{\pi\, \omega^2}+{\rm local\, terms}\,.
\eea

We note that the first two contributions listed in  \eqref{eq:M-limits} approach $E$ and $m$ dependent constants at late time. The last contribution coming purely from the non-zero energy sector yields,
upon analytic continuation and using the replica formula, the following behaviour of the quantum expectation value of length in the leading order of expansion in $\omega$
\bea
\label{eq:late-time-complexity-2}
&&\langle \ell(t)\rangle_{m}=C_{0,m}-\frac{e^{S_0}\sin( 2\pi m)}{Z^{l}_m(\beta)}
 \bigg(g(m)
+4\sqrt{\pi}
\int_{m^2}^\infty dE\;e^{-\frac{1}{2}\beta E}\;\frac{{\hat \rho}_{0,m}(E) \cos tE}
{\cosh{2\pi \sqrt{E - m^2}-\cos{2 \pi m}}}
\bigg)
\cr \nonumber\\
&&\;\;\;\;\;\;\;\;\;\;\;\;\;\;\;\;\;\;\;- \frac{2 e^{S_0}}{\pi Z^{l}_m(\beta)} \int_{m^2}^\infty dE \, e^{-\beta E} \sqrt{E-m^2}\, {\hat \rho}_{0,m}(E) \int_{-\infty}^{\infty}\frac{e^{-i \omega t}}{\omega^2}  ,
\eea
where $C_0$ is a time independent term coming from the local terms and 
$g(m) = \psi ^{(0)}(-2 m)+\psi ^{(0)}(2 m+1)+\log (4)$. From the last term of the
equation
\eqref{eq:late-time-complexity-2}  one gets  a linear growth for matrix elements of the complexity at late times, $\langle \ell(t)\rangle_{m} \sim t$.  Note that due to 
particulare stracture of the wavefunction we have $\delta_{mm'}$ for the  matrix elements of the complexity and therefore in the above equation we have dropped 
$m'$ in the expression. Of course at the end of the case one needs to sum over $m$.

It is also interesting to note from the expression \eqref{eq:late-time-complexity-2} that, although the zero modes contribute in a constant at late times, their convolutions with higher modes give, rather, a mild oscillating time dependence to the late-time behaviour of complexity.

So far we have only computed complexity for left hand side. However, due to the symmetric
structure of the wavefunction, the final result applies for right hand side as well.
Of course, when we compute to the complexity of TFD state, due to convolution of left and right hand sides one gets extra contribution as we will demonstrate in the
following. In the TFD case, in order to employ the modified replica trick \eqref{eq:replica-formula}, one essentially needs to compute the matrix element
\be
\langle y^{2N}\rangle_{u,mm'}= \frac{1}{Z^{\rm TFD}(\beta)}\int_0^\infty \frac{dy}{y}\int_0^{2\pi q}\frac{da}{2\pi q}\int
d\chi_1 d\chi_2\;
\bar \Psi_{D,m'}^{\rm TFD}(u,y,a)\Psi_{D,m'}^{\rm TFD} (\beta-u,y,a)\left(\frac{y}{4}\right)^{2N}
\ee

where
\be
\Psi_{D,m}^{\rm TFD}(\beta, y, a) = e^{S_0/2}  e^{2ima} \,
\hat f^{\rm TFD}_m(y,a), 
\ee
with $f^{\rm TFD}_m(y,a)$ given by \eqref{eq:TFD2}. Writing explicitly in terms of the mode functions \eqref{eq:TFD2}, this yields\footnote{Similar matrix elements in the context of the computation of the two-point function is also presented in \cite{Lin:2022zxd}.}
\bea
\label{eq:replicayNTFD-1}
&&\langle y(u)^{2N}\rangle_u=\frac{e^{S_0}}{Z^{\rm TFD}(\beta)}\bigg\{\frac{4}{{\pi}}
\sin^2( 2\pi m) \, {\cal M}^{(1)}_{mm'}(N)
\\ 
&&\hspace{1.8cm}-\frac{2}{\sqrt{\pi}}
\sin( 2\pi m)\int_{m^2}^\infty dE_2 e^{-(\beta-u) E_2} 
{\hat \rho}_{0,m}(E_2)\, {\cal M}^{(2)}_{mm'}(N, E_2)
\nonumber\\
&&\hspace{1.8cm}-\frac{2}{\sqrt{\pi}}
\sin( 2\pi m')\int_{m'^2}^\infty dE_2 e^{-u E'_1} 
{\hat \rho}_{0,m'}(E'_1)\, {\cal M}^{(2')}_{mm'}(N, E'_1)
\nonumber\\
&&\hspace{1.8cm}+\int_{m^2}^\infty dE_1 dE_2\;e^{-u E_1-(\beta-u) E_2} \, {\hat \rho}_{0,m}(E_1){\hat \rho}_{0,m}(E_2) {\cal M}^{(3)}_{mm'}(N, E_1, E_2) 
\nonumber\\
&&\hspace{1.8cm}+\int_{m'^2}^\infty dE'_1 \int_{m^2}^\infty dE_2\;e^{-u E'_1-(\beta-u) E_2} \, {\hat \rho}_{0,m'}(E'_1){\hat \rho}_{0,m}(E_2) {\cal M}^{(3')}_{mm'}(N, E'_1, E_2) \bigg\},\nonumber
\eea
where $E'_{i}= k_i^2 + m'^2$, and
\bea
&&{\cal M}^{(1)}_{mm'}(N) = 2^{2N - 3} \frac{N\Gamma^2(N)}{\Gamma(2N)}\Gamma(N-2m)\Gamma(N+1+2m) \delta_{m,m'} \nonumber \\
&&{\cal M}^{(2)}_{mm'}(N, E_2) = 2^{2N-2} \frac{|\Gamma(N+m+ik_2)\Gamma(N-m+ik_2)|^2}{\pi \, \Gamma(2N)} \nonumber \\ 
&&\hspace{3cm} \times\left[(E_2+N(N+2m)) \delta_{m',m} + (E_2+N(N-2m)) \delta_{m',m-\frac{1}{2}}\right]\nonumber \\
&&{\cal M}^{(2')}_{mm'}(N, E'_1) = 2^{2N-2} \frac{|\Gamma(N+m'+ik_1)\Gamma(N-m'+ik_1)|^2}{\pi \, \Gamma(2N)} \nonumber \\ 
&&\hspace{3cm} \times\left[(E'_1+N(N+2m')) \delta_{m',m} + (E'_1+N(N-2m')) \delta_{m',m+\frac{1}{2}}\right]\nonumber \\
&&{\cal M}^{(3)}_{mm'}(N, E'_1, E_2) =  2^{2N-1}\,(E_1+E_2+N^2)
\frac{|\Gamma(N+i(k_1+k_2)) \Gamma(N+i(k_1-k_2))|^2}{\pi^2 \,\Gamma(2N)} \delta_{m,m'} \nonumber \\
&&{\cal M}^{(3')}_{mm'}(N, E'_1, E_2) =  2^{2N-1}\,
\frac{|\Gamma(\frac{1}{2}+N+i(k_1+k_2)) \Gamma(\frac{1}{2}+N+i(k_1-k_2))|^2}{\pi^2 \,\Gamma(2N)} \nonumber \\
&&\hspace{3.2cm}\times \left(\delta_{m', m-\frac{1}{2}} + \delta_{m', m + \frac{1}{2}}\right).
\eea
We can now proceed to evaluate the late time behaviour of complexity following the same procedure as adopted for the computation involving the left side. This involves making the
analytic continuation $u=\frac{\beta}{2}+it$ and plugging the resulting expression in the replica formula.
Again, since we are interested particularly in the late time behaviour, we need to consider the coincident limit. However, we note here that unlike the previously considered cases, the matrices ${\cal M}^{(2')}_{mm'}(N, E'_1)$ and ${\cal M}^{(3')}_{mm'}(N, E_1, E_2)$ have non-vanishing off-diagonal components. Therefore, it is illuminating to use a slightly different set of variables
\be
k^2 =\frac{k_1^2+k_2^2}{2}\, , \qquad \omega =k_1^2 -k_2^2.
\ee
Using this notation and in the limit of $\omega \rightarrow 0$, one gets
\bea
\label{eq:M-limits-TFD}
&&\lim_{N\rightarrow 0}\partial_N {\cal M}^{(1)}_{mm'}(N)
\approx -\frac{\pi}{4 \sin{2 \pi m}} \left(\psi ^{(0)}(-2 m)+\psi ^{(0)}(2 m+1)+\log (4)\right) \delta_{m,m'}\nonumber \\&&
\lim_{N\rightarrow 0}\partial_N {\cal M}^{(2)}_{mm'}(N, E_2)
\approx \frac{\pi}{\cosh{2\pi k -\cos{2 \pi m}}}\left(\delta_{m',m} +\delta_{m',m-\frac{1}{2}}\right)+{\rm local\, terms}\,\nonumber \\&&
\lim_{N\rightarrow 0}\partial_N {\cal M}^{(2')}_{mm'}(N, E'_1)
\approx \frac{\pi}{\cosh{2\pi k -\cos{2 \pi m'}}}\left(\delta_{m',m} +\delta_{m',m +\frac{1}{2}}\right)+{\rm local\, terms}\,\nonumber \\&&
\lim_{N\rightarrow 0}\partial_N {\cal M}^{(3)}_m(N,E'_1,E'_2)
\approx \frac{1}{{\hat \rho}_{0,m}(E)}\,\frac{2 k}{\pi\, \omega^2} \, \delta_{m,m'} +{\rm local\, terms}\, \nonumber \\&&
\lim_{N\rightarrow 0}\partial_N {\cal M}^{(3')}_m(N,E'_1,E'_2)
\approx \frac{1}{\cosh{2\pi k}} \left(\delta_{m', m-\frac{1}{2}} + \delta_{m', m + \frac{1}{2}}\right)+{\rm local\, terms}\,.
\eea
Finally taking all terms together and summing over $m$ and $m'$  one arrives at
\bea
\label{eq:late-time-complexity-TFD-2}
&&\langle \ell(t)\rangle = \sum_m \sum_{m'} \langle \ell(u)\rangle_{mm'}
 =  C_{0}-\frac{e^{S_0}}{Z^{\rm TFD}(\beta)}\sum_{0\leq m\leq\frac{1}{2}} \sum_{m'}\sin( 2\pi m)
 \bigg[g(m)  \, \delta_{m,m'} \\ 
 &&\hspace*{3cm} + 8\sqrt{\pi} \int_{m^2}^\infty dE\;e^{-\frac{1}{2}\beta E}\;{\hat \rho}_{0,m}(E) \frac{\cos tE  \left(\delta_{m',m} +\delta_{m',m-\frac{1}{2}}\right)}
{\cosh{2\pi \sqrt{E - m^2}-\cos{2 \pi m}}}\bigg]\nonumber \\
&&\hspace*{3cm}- \frac{2 e^{S_0}}{\pi Z^{\rm TFD}(\beta)} \sum_{m}\int_{m^2}^\infty dE \, e^{-\beta E} \sqrt{E-m^2}\, {\hat \rho}_{0,m}(E) \, \int_{-\infty}^{\infty}\frac{e^{-i \omega t}}{\omega^2}\,,  \nonumber
\eea
where $C_0$ comes from the local terms which we do not explicitly write down. 
Note that it has the same structure as that of the left hand side which exhibits linear growth at late times. 

\subsection{Comment on the saturation of complexity}
\label{sec:comment-saturation}
In the previous section we have computed complexity for ${\cal N}=2$ model where
we have found the linear growth at late times. However, we also expect that the complexity must 
saturate at a time of the order of $t\sim e^{S_0}$ which could not be achieved from our 
disk level computation. Of course, this is expected from the intuition we gain from our previous computations backed by \cite{Alishahiha:2022kzc} and \cite{Iliesiu:2021ari}. It is legitimate to assume that in order to see the late-time saturation, one does need to take into account the higher genus contributions to the complexity as well.
The latter requires the knowledge of the trumpet partition function as well as the non-perturbative 
correction to the spectral correlation
$\langle\rho(E_1)\rho(E_2)\rangle$ in the coincident limit.

Sadly, we do not precisely know the partition function for 
the trumpets for the supersymmetric JT gravity with ${\cal N} = 2$ supersymmetry. Second, for such a supersymmetric case, we also do not have an expression for the spectral correlation function. Therefore in general it is not straightforward to 
see the saturation period in the ${\cal N}=2$ model. However, from our computation for the disk level, we can extract two interesting observations which might possibly provide some intuitions regarding what one might ideally expect for the late time growth of complexity of these systems. This subsection should be considered as a speculative aside.

To proceed, we note that the state with the highest  component of ${\cal N}=2$ multiplet is obtained by
setting $j_r+j_l=1$  in \eqref{HCom} and the corresponding wavefunction can be written as
\be\label{Dwave}
\Phi^{h}_{D,m}(\beta,y,a)=e^{S_0/2}e^{2ima}\int_0^\infty dk\; e^{-\beta E}\;\frac{4k\sinh2\pi k}{\pi^2}
K_{2ik}(y)\,.
\ee
Due to the $R$ charge of the above wavefunction, it has non-zero fermion number.
The partition function associated with this wavefunction is
\be
\label{eq:partition-highest}
Z_{D,m}^h(\beta)=e^{S_0}\int_0^\infty dk\;e^{-\beta E}\;\frac{k\sinh 2\pi k}{\pi^2}\,.
\ee
Staring at this partition function and the wavefunction, one observes that it has almost
the same form as those of the bosonic JT gravity, with the only difference that, in this case, there is shift in the energy. Of course, 
this shift implies a non-trivial low-energy physics arising out of a non-zero density
of state (degeneracy) at zero energy. The density of state of the zero modes can be read off from the disk partition function \eqref{eq:partition-highest} and takes the form 
\be
\rho_0 =\frac{\sin 2\pi m}{2\pi^2}.
\ee

With this, we can rewrite the expressions for the wavefunction and the corresponding partition function incorporating the zero modes explicitly. 
\be
Z_{D,m}^h(\beta)=e^{S_0}\left(-m\sin(2\pi m)+\int_0^\infty dk\;e^{-\beta E}\;\frac{k\sinh 2\pi k}{\pi^2}\right)\,.
\ee
The wavefunction
for the other elements of the multiplet can be obtained by acting on the above wavefunction
with $\bar Q_{l,r}$, which essentially results in the equation \eqref{eq:phi-mode-exp}. 

Noting the similarities of the density state and the wavefunction for non-zero modes of ${\cal N}=2$ case with those of the bosonic JT gravity, it is tempting to propose the following expression for the trumpet partition function for the highest component of the ${\cal N}=2$ multiplet 
\be
Z^h_{T,m}(\beta,b)\sim \cosh(bm)+\int_0^\infty dk\;e^{-\beta E}\cos(bk)\,.
\ee
It is then straightforward to compute the Hartel-Hawking wavefunction for a trumpet pinched off by the disk region given by the wavefunction \eqref{Dwave} as
\be
\Phi_{T,m}^h(\beta,b,a,y)\sim e^{-S_0/2}e^{2ima}\left(\cosh(bm)K_{2m}(y) +
\int_0^\infty dk\;e^{-\beta E}\cos(bk) 
K_{2ik}(y)\right)\,.
\ee
Using the supersymmetric charges, one can also find the corresponding wavefunction for the other component of multiplet. 
\be\label{HHwT}
\Phi_{T}^{l,r}(\beta,b,z) \sim e^{-S_0/2}e^{2ima}\left(\cosh(b m)\hat f_{m}^{l,r}(y,a) +
\int_0^\infty dk\;e^{-\beta E}\;\frac{\cos(bk)}{E} 
\hat f_k^{l,r}(y,a)\right)\,,
\ee
where $\hat f$'s  are defined in \eqref{eq:f-lr}. The above expression can be justified from the following observation. From the equation \eqref{eq:late-time-complexity-2}, it is clear that as far as the late time behavior of complexity is concerned, the relevant part in the partition function is
 \bea
\label{eq:partition-nonzero}
&& Z^{l,r}_{m,m'}(\beta)=e^{S_0}
\int_0^\infty dk \, e^{-\beta E}\;
\frac{k\sinh 2\pi k}{E}\; \delta_{mm'}\, \nonumber \\
&&\hspace{1.7cm}=2\pi^2\int_0^\infty d\xi\,e^{-(\beta+\xi) m^2}\left(e^{S_0}
\int_0^\infty dk \, e^{-(\beta+\xi) k^2}\;
\frac{k\sinh 2\pi k}{2\pi^2}\right)\; \delta_{mm'}\, \nonumber \\
&&\hspace{1.7cm} = 2\pi^2\int_0^\infty d\xi\,e^{-(\beta+\xi) m^2}
Z_{D,0}(\beta+\xi)
\eea
 where the $Z_{D,0}(\beta+\xi)$ is the disk partition function of non-supersymmetric JT gravity with an effective regularized length $\beta+\xi$ of the asymptotic boundary.
 Indeed it is very much reminiscent of the disk partition function for non-supersymmetric JT gravity models except for the additional integration over $\xi$. Then
 one might propose an effective trumpet partition function as follows
\bea
&&Z^{l,r}_{T,m,m'}(\beta,b)= 2\pi^2\int_0^\infty d\xi\,e^{-(\beta+\xi) m^2}
Z_{T,0}(\beta+\xi,b)
\, \nonumber \\
&&\hspace{2.3cm}=2\pi^2\int_0^\infty d\xi\,e^{-(\beta+\xi) m^2}\left(\frac{1}{\pi}
\int_0^\infty dk \, e^{-(\beta+\xi) k^2}\;
\cos(bk)\right)\; \delta_{mm'}\, \nonumber \\
&&\hspace{2.3cm} =2\pi\int_0^\infty dk \, e^{-\beta E}\;
\frac{\cos b k}{E}\; \delta_{mm'}
 \eea
Here, as it is indicated, $Z_{T,0}(\beta+\xi,b)$ is the trumpet partition function 
of non-supersymmetric JT gravity with the effective regularized length $\beta+\xi$ of the asymptotic boundary. It is easy to check that this part of the trumpet partition function can  be obtained directly from  \eqref{HHwT}.
 
To conclude, we note that, looking at  the late time behavior of complexity as in \eqref{eq:late-time-complexity-2} one observes that the late time it is entirely governed by the non-zero energy sector. The contributions coming from the zero-energy supersymmetric sectors do not lead to linear growth of complexity at late times. The behavior is, indeed, 
very similar to that in the ${\cal N} = 0$ case without supersymmetry. Such a similarity is expected since there is a large gap in the spectrum between the zero energy and the higher energy sectors. Using the same intuition, it is therefore very natural to expect, even beyond the disk level this very feature of the late time growth of complexity would prevail. 

Using the  proposed wavefunction \eqref{HHwT} (let's say for left hand side) one can compute higher genus corrections to the complexity. Indeed, as far as the late time behavior is concerned, we get
\be
\langle \ell(t)\rangle_m \sim 
- \frac{ e^{-S_0}}{ Z(\beta)} \int_{m^2}^\infty dE \, e^{-\beta E} \frac{\sqrt{E-m^2}}{ {\hat \rho}_{0,m}(E)} \, \int_{-\infty}^{\infty}\frac{e^{-i \omega t}}{\omega^2}
\left\langle \rho\left(E+\frac{\omega}{2}\right) \rho\left(E-\frac{\omega}{2}\right)
\right\rangle_{\omega\rightarrow 0},
\ee
If we further use the fact that at late times the short range correlations give the main contribution to the connected part of the spectral correlation function which, in turn, yields the 
universal sine kernel term \eqref{eq:sine-kernel}, one has
\be 
\left\langle \rho\left(E+\frac{\omega}{2}\right) \rho\left(E-\frac{\omega}{2}\right)
\right\rangle_{\omega\rightarrow 0}\approx e^{2S_0}{\hat \rho}_{0,m}^2(E)-
 \frac{\sin^2(e^{S_0}{\hat \rho}_{0,m}(E)\omega^2)}{\omega^2}.
 \ee
 This sine kernel results in the saturation of the late time growth of complexity, 
in a spirit similar to the that obtained in the non-supersymmetric case \cite{Alishahiha:2022kzc}.


\subsection{Matter correlation functions} 
As we have already discussed before, using the wavefunction formalism one can compute the full quantum gravity expressions for the matter correlation functions in 
JT gravity coupled to matter fields. It is also expected the same procedure should go through for the ${\cal N} = 2$ supersymmetric case as well. The basic procedure remains the same, namely, 
to construct a certain Kernel which in turn can be used to dress the quantum field theory correlation functions to produce gravity correlators.

As before, the Kernel essentially should be read off from the Hartle-Hawking wavefunctions we considered in the previous section. In what follows, inspired from our previous computations, we present the form of the ${\cal N} = 2$ supersymmetric kernel. 
In the present case, one notice that the near boundary correlators can be 
labeled by both bosonic and fermionic coordinates $( x,\theta_-,\bar \theta_-)$. With this, the near boundary correlators assume the following form
\be
\prod_i \xi_i^{\Delta_i}e^{i\alpha_i a}\langle {\cal O}_1(x_1,\theta_{1-},\bar \theta_{1-})\cdots{\cal O}_n(x_n,\theta_{2-},\bar \theta_{2-}) \rangle\,.
\ee
The corresponding supersymmetric kernels are functions of both 
bosonic and fermionic variables that we collectively denote as
by ${\bf x}=(\xi, x, a,\theta_-,\bar \theta_-)$. To write down the kernels, one needs to provide a map between the wavefuction variables to the variables, ${\bf x}$ above. In this case the map takes the form
\be
\beta\rightarrow u,\;\;\;\;y\rightarrow \frac{4}{\omega},\;\;\;\;
a\rightarrow \sigma,\;\;\;\;\chi_{1,2}\rightarrow \eta_{1,2},
\ee
where
\cite{Lin:2022zxd}
 \bea\label{dif}
&& \omega = \frac{ x_1 - x_2  - \theta_{1-} \bar \theta_{2-} - \bar \theta_{1-} \theta_{2-} }{\sqrt{\xi_1\xi_2}}\,,\;\;\;\;\;\;\;\;\;\;
 e^{i\sigma}  =  e^{i ( a_1 - a_2) } \left( 1 - { ( \theta_{1-} - \theta_{2-}) ( \bar \theta_{1-} - \bar \theta_{2-}) \over (x_1 - x_2) }  \right),
 \nonumber\\  
  &&\eta_1 = \chi_1 -e^{ -i a_1}  { (\bar \theta_{1- } - \bar \theta_{2-} ) \over \omega\sqrt{\xi_2} \,},\;\;\;\;\;\;\;\;\;\;\;\;\;\;\;\;\;
 \eta_2 =  \chi_2 - e^{- i a_2} { (\bar \theta_{1- } - \bar \theta_{2-} ) \over\omega \sqrt{\xi_1} }\,. 
 \eea
Using 
this notation, up to a phase, we propose 
\bea
&&K^{l,r}_m(u,{\bf x}_1,{\bf x}_2)=e^{S_0/2}e^{2im\sigma}\int_0^\infty
dk e^{-uE}\;\frac{2k\sinh 2\pi k}{\pi E}\,\hat f^{l,r}_k(4/\omega,\sigma),\cr &&\cr 
&&K^{0}_m({\bf x}_1,{\bf x}_2)=-\frac{2}{\sqrt{\pi}}
e^{S_0/2}e^{2im\sigma}\sin(2\pi m)
\hat f^l_{m}(4/\omega,\sigma)\,.
\eea
Note that, in our convention, the kernels are defined for ordered points $x_1>\cdots >
x_n$ and for $x_i>x_j$ one has $K(u,x_j,x_i)=\bar K(u,x_j,x_i)$. It worth noting that the kernels also acquire a phase that are in the form of $e^\phi$ with
\be
\phi =   \gamma_1 - \gamma_2 + \frac{ \xi_1+\xi_2}{ \omega\sqrt{\xi_1\xi_2} } +2 { ( \frac{\chi_1 e^{  i a_1 }}{\sqrt{\xi_2}} -  \frac{\chi_2 e^{  i a_2 }}{\sqrt{\xi_1}}) }{(\theta_{1-} - \theta_{2-})   \over \omega } +
  {(\xi_1-\xi_2)(\theta_{1-} - \theta_{2-})  (\bar \theta_{1-}- \bar \theta_{2-})   \over \omega^2\xi_1\xi_1} \nonumber
\ee
To understand the parameters appearing in the definition of the phase and those in 
\eqref{dif}, one notes that the elements of ${\cal N}=2$ supergroup $SU(1,1|1)$ may
be parametrized as follows
\be
g=e^{-xL_-}e^{\theta_- G_-+\bar\theta_- \bar G_-}e^{\rho L_0}
e^{\theta_+ G_++\bar\theta_+ \bar G_+}e^{\gamma L_+}e^{iaJ}
\ee
Then in the $m$-th sector, the gravity correlation function is obtained as
\bea
\label{eq:grav-corr-N2}
&&\langle {\cal O}_1(u_1)\cdots {\cal O}_n(u_n)\rangle^{l,r}_m=
\int_{x_1>\cdots >x_n}\frac{\prod_i d\mu_i}{{\rm Vol}(SU(1,1|1))}
K^{l,r}_m(u_1,{\bf x}_1,{\bf x}_2)\cdots K^{l,r}_m(u_n,{\bf x}_n,{\bf x}_1)\cr &&
\cr&&\hspace*{8cm}\times
\prod_i\xi_i^{\Delta_i}e^{i\alpha_ia_i}\langle {\cal O}_1(x_1,\theta_1)\cdots
 {\cal O}_n(x_n,\theta_n)\rangle \nonumber\\
 \eea
A similar expression can be written for zero energy modes as well, though in this case they are independent of $u_i$'s. Here the integration measure is defined as $d\mu_i=d\xi_i dx_i d\theta_{i-}d\bar\theta_{i-}da_id\eta_i$.

Using \eqref{eq:grav-corr-N2} one could compute correlation function of matter fields. The way we have constructed the kernel, it is clear that resultant correlation functions should match those computed from wavefunction formalism. There is a word of caution - strictly speaking, the correlation function we compute this way is is only an on-shell correlation function, in the sense that, in principle, the kernel could also have a fermionic
term which, during the actual computation, we have set to zero, through a gauge fixing.

\section{Conclusion and Outlook}
In this work we showed that the modified replica method proposed in \cite{Alishahiha:2022kzc} can be used as a useful trick to extract the late-time behaviour of complexity for supersymmetric JT gravity models as well. One distinct advantage of this method is that it does not require an explicit averaging over infinite number of geodesics in contrary to the definition of complexity adopted in \cite{Iliesiu:2021ari}. In this work, we considered two classes of supersymmetric JT gravity models - JT gravity theories with ${\cal N} = 1$ and ${\cal N} = 2$ supersymmetries respectively. While for the ${\cal N} = 1$ theory, we are able to achieve the late time saturation of complexity as in the bosonic case, for the ${\cal N} = 2$ case, it turned out hard to offer a precise computation beyond the disk level due to the lack of knowledge of the exact trumpet wavefunction and full the spectral correlation including the connected pieces. At the disk level, as expected, it shows linear growth. However, using some educated guesses based on the spectrum of these theories, in particular, the large gap near the zero energy spectrum, we are able to come up with an intuitive understanding of how complexity should behave at late time if we also include the trumpet geometries. Interestingly, one can argue, with reasonable assumptions for the spectral correlation, that at times ${\cal O}(e^{S_0})$, we should still expect a saturation in the growth of complexity.
It will nevertheless be very interesting if one can make this computation more precise with less assumptions. In particular, it would be extremely illuminating to investigate, if our proposed expression for trumpet partition function can be verified by direct computations as that done in \cite{Stanford:2017thb}. While this is a challenging task, it might provide the fundamental insight on the universality of the JT models, beyond the disk level. We postpone a rigorous study in this direction for a future work.

Another interesting future direction of work would be to understand better the crosscap states \cite{Yan:2022nod} we briefly mentioned in this work and computed the correlation functions for. These crosscap states result in the non-decaying behaviour of matter correlation functions, which we confirmed through our computation. It would be extremely interesting to understand these states better and compute the contributions thereof in the late-time behaviour of matter correlation functions and complexity upon including the contributions coming from higher topologies. 

\acknowledgments
We would like to thank  Johanna Erdmenger and Kyriakos Papadodimas for useful discussions. M.A. would also like
to thank  Department of Theoretical Physics of CERN for very warm hopitality.

\bibliographystyle{JHEP}
\bibliography{literature}

\end{document}